# Survey on Internet of Things enabled by 6G Wireless Networks

[a]Sridhar Iyer, [b]Rahul Jashvantbhai Pandya, [c]Rakhee Kallimani, [d]Krishna Pai, [e]Rajashri Khanai, [f]Dattaprasad Torse, [g]Swati Mavinkattimath

[a,c,d,e,g]Department of ECE, KLE Dr. M.S. Sheshgiri College of Engineering and Technology, Udyambag, Belagavi, KA, India- 590008.
e-mail: sridhariyer1983@klescet.ac.in; rakhee.kallimani@klescet.ac.in; krishnapai@klescet.ac.in; rajashrikhanai@klescet.ac.in; swatim@klescet.ac.in

[b]Department of EE, Indian Institute of Technology, Dharwad, WALMI Campus, PB Road, KA, India-580011.
e-mail: rpandya@iitdh.ac.in

[f]Department of CSE, KLE Dr. M.S. Sheshgiri College of Engineering and Technology, Udyambag, Belagavi, KA, India- 590008.
e-mail: datorse@klescet.ac.in

**Abstract**

The 6G wireless technology is visualized to revolutionize multiple customer services with the Internet of Things (IoT), thereby contributing to a ubiquitous intelligent society comprising autonomous systems. In this chapter, we conduct a detailed survey on the IoT networks with 6G wireless networks and investigate the trending possibilities provided by the 6G technology within the IoT networks and the related utilization; Firstly, we detail the breakthrough IoT technologies and the technological drivers which are anticipated to strengthen IoT networks in future. Next, we present the relevant use cases detailing the discussion on the role of the 6G technology within a broad spectrum of IoT potential applications. Lastly, we highlight the several research scope and challenges and list the potential research needs and encourage further research within the thrust area of IoT enabled by 6G networks.

# 1. Introduction

In recent years, the potential growth in IoT is due to the technological advancements in wireless communications and smart device technologies with extensive potential in sensing and computing facilities, thereby interconnecting multiple physical objects to the Internet [1]. As an enabler, IoT ensures persistent communications and self-management among the several diverse devices without the need of intervention of human, that is potentially a revolutionary change within the industries. It also benefits society via completely intelligent and self-regulating systems. It has been predicted, in coming future the IoT will demonstrate tremendous development. E.g., according to Cisco, by 2030 it is expected to connect approximately 500 billion devices, as compared to 26 billion devices in 2020 [2] to internet. Further, IHS Markit has analysed a rise in the rate of global connected devices by 12 % annually, that is approximately 27 billion devices to 125 billion devices over a period from 2017 to 2030 respectively [3]. Furthermore, Globe Newswire has forecasted the growth in 5G -IoT network market from approximately $ 694.0 million to $6,285.5 million over a period of 5 years from 2020 [4]. The connections of tiny things (Nano-Things) to internet has also gained prominent attention in building the advanced ecosystems [5]. Hence, currently, IoT has become a key attribute and has received much attention from both the industry and the research community owing to the capability of delivering quality services to the modern customers in many aspects of trending lifestyles [6].

Within the IoT, the Internet is an enabler that supports the IoT networks and the applications, and in this regard, the first to the fifth generations of mobile technology are been proposed and commercially deployed. Specifically, the latest 5G wireless technology, various enhanced services like massive Machine-Type Communication (mMTC), the enhanced Mobile Broadband (eMBB), and Ultra-Reliable and Low-Latency Communication (URLLC) have been proposed, which will be able to provide various service opportunities to the ecosystems of IoT for better performance in terms of throughput(high), latency(low) and energy-efficiency [7-8]. Within the 5G networks, researchers have been investigating new methods of communicating the connecting devices in sophisticated applications viz., Wireless Brain-Computer Interfaces (WBCI), five sense communications, holographic communications, etc., that will result in truly immersive experience into an isolated environment. Simultaneously, advances within personal communications promote the evolution of smart verticals within the 5G technology to a higher level, including healthcare, remote education and training, Industry Internet, fully autonomous driving, and super smart homes and cities, etc. Overall, during this paradigm shift, the connection of physical and cyberspace of the communication systems, it is noted that IoT plays a prime role in enabling the aforementioned trending applications.

However, with the rapidly expanding IoT networks and the emergence of multiple smart devices, 5G technology will face the daunting challenge of supporting the next generation applications with increased technical requirements such as completely dynamic and intelligent services, autonomous devices, unmanned aerial vehicles, etc. In specific, the next-generation IoT enabled applications will (i) requisite superior performances in regard to the data rate, latency, coverage, and localization, (ii) be more data- and computation-intensive, which will exceed the URLLC and mMTC range of the 5G technology [7], (iii) face difficulty in efficiently managing the massive IoT devices, and (iv) face massive generated data which will be accompanied with serious security and privacy issues [8]. Hence, with the rapid evolution of the IoTs, the 5G technology will met to the limitations gradually and will not be able to provide any assist to the majority of the recent modern applications. Hence, a strong motivation exists to enable the massive IoTs to extend the 5G technology capabilities towards the next-generation sixth (6G) generation networks.

In view of the aforementioned, the research community has already begun conceptualizing the 6G wireless network architecture [9, 10] and has made promising progress towards the concept of IoT with 6G to meet the needs of the future ubiquitous smart society. The major applications of 6G-IoT will include healthcare IoT, vehicular IoT, and autonomous driving, satellite IoT, industrial IoT, and many more, requiring quality of service and enhanced experience of users from the systems which can be provisioned by the 6G technology owing to the superior features such as communications with ultra-low latency, throughput to be extremely high, customer services related to satellite-based, autonomous massive networks [11-13]. Further, the research community has already identified that the capacity levels required by the next-generation applications will be outstanding, in turn, increase the applications and the organization of 6G enabled IoT networks for sensing of data, connection of devices, wireless- communication, and management of network. Hence, recently, after having identified the immense potential of the 6G-IoT, much effort has been conducted in research within this promising area [10, 14-16]. During these efforts, it has been identified that the 6G technology, owing to the special features and enhanced capabilities, will be a crucial factor to support the future generation of IoT networks and applications as it will provide complete spatial wireless coverage with integration of all the functionalities ranging from sensing, computing, communication, to smart and self-organized autonomous control. Consequently, compared to 5G networks, the 6G networks have been anticipated to contribute extensive range and improved expandability to promote connectivity of IoT devices and deliver quality service [17].

## 2. Breakthrough IoT Technologies

The key technology is to integrate variant devices with wireless communication systems, IoT targets at connecting various devices(things) to the Internet, thereby developing an environment connected; wherein the sensing of data, computations, and communications are carried out without any interventions of human automatically. In recent years, with the expanse of human activities towards extreme environments such as higher altitudes, outer space, under-sea/ocean, it is required to build a ubiquitous Internet of Everything (IoE) which is omniscient and omnipotent, and which can recognize the connection at any moment and in any place with differing needs. In order to accomplish the aforementioned, a four-tier network architecture, reinforced by computing at edge, will be regulated by the 6G technology [18]. Specifically, edge computing has been proposed as a crucial factor [19, 20]. From the users' context, edge computing aids IoT devices perform the jobs directly, outperforming the centralized cloud computing technology due to the low latency and distributed nature. From the context of system, Machine Learning (ML) enabled edge intelligence, is ensured via computing at edge, which helps handle IoT enabled systems through multiple intelligent methods.

6G will enhance multiple breakthrough technologies as an omnipotent network, including ML and Blockchain. As an intelligence enabling technology, ML is extensively employed in various forms of IoT-enabled applications, ranged from the application and network layer to the perception layer. Recently, in wireless systems, ML is been adopted to deal the related challenges/issues and provide an organized way for future extensive IoT communications [21-23]. Regarding 6G system in IoT networking, ML algorithms will be extensively employed to solve multiple resource allocation, power allocation, transmit scheduling, traffic offloading, etc. [24].

Further, the current IoT network models are centralized, with cloud server for connection and IoT devices connected to an individual gateway for transferring the data. However, the aforementioned setup is not suited for the forthcoming IoT devices and the amount of data they will be sharing owing to (i) higher price of the centralized cloud maintenance charges and networking equipment, (ii) interoperability is low on account of the bounded data exchange with the other centralized infrastructures, and (iii) issues/challenges related to security due to the untrustworthy single gateway and centralized cloud server. In this regard, recently, Blockchain, a decentralized distributed ledger, has been proposed as a key solver to multiple problems which are faced by the current models, including security improvement [25, 26]. Hence, in addition to intelligence via ML, dynamic network management will be enabled by blockchain technology with decentralization and low cost [27, 28].

Also, along with intelligence in the 6G systems, the ML functions will be extendible to the network edge mainly due to the computational capabilities of the edge nodes [29]. The aforementioned is the novel paradigm known as edge intelligence [30, 31], which uses the convergence provided by ML along with edge computation. Also, lately, Federated Learning (FL) emerged as a distributed collaborative ML technique that is transforming the edge/fog intelligence architectures [32]. Another recent technology that has received much attention from the research community is Reconfigurable Intelligent Surfaces (RISs) [33]. In the RIS technology, there exist artificial planar structures. Electronic circuits enable every element in a software-defined manner by reflecting the impinging electromagnetic wave [34]. RIS is considered as a potential solution for supporting the design of wireless system owing to the configurability characteristic, which facilitates propagation of signal, modelling of channel, and acquisition, thereby ensuring benefits of smart radio environments for the 6G enabled applications. Further, the multiple diverse applications via 6G-IoT will require very large bandwidth and very low latency, for which Tera-Hertz (THz) communications have been envisioned as a driving technology [35]. A benefit of the THz spectrum is that it can deal with issues related to scarcity of spectrum in wireless communications and can predominately improve the capacities of wireless system within the 6G-IoT. Owing to the aforementioned key benefits of the THz communications, various study related to IoT have been conducted in regard to the 6G technology [36, 37].

Lastly, in line with the vision of IoE, 6G will aim to connect all parts of the planet viz., land, sky, space, underwater, and to achieve this aim, 6G-IoT will require a unified communication platform. Further, in view of achieving a broad coverage and ubiquitous connectivity to support the various IoT applications, a cell-free (de-cellular) and 4-tier (space, air, terrestrial, and underwater) large-dimensional communication network will be necessary [10]. The 4-tiers will comprise of:

i. **Space Communication** will use various satellite types to provide wireless coverage for the terrestrial networks' areas. Further, 6G technology will introduce the concept of satellite-terrestrial communications [38].
ii. **Air Communication** will introduce the concept of base stations in the air with flying base stations constructed by UAVs and balloons to cover expanded coverage in remote areas [39].
iii. **Terrestrial communication** which will aim to provide coverage and connectivity over ground using the base stations, user equipments, servers, etc., which will be interconnected and use the THz band in view of large bandwidth [40].
iv. **Underwater Communication** will provide connectivity services to underwater IoT devices [41].

## 3. Technological Drivers

In recent years, IoT enabled by 5G technology has been transforming and generating the Industrial Revolution 4.0 impacting every sector of human society inclusive of Smart- healthcare, Smart-education, Industry Internet, Autonomous Driving, Smart Homes, and Smart Cities. Simultaneously, multiple IoT enabled applications have emerged creating new phase in technological progress. These evolving IoT applications impose stricter and robust requirements on wireless networks. In the present section, we compile the key technological drivers resulting in the fast-paced evolution of the IoT towards the 6G-IoT.

1) **Holographic communications**: The initial novel form of interactions known as holographic communications, manipulates motion 3D images and projects in real-time. The technology here, captures the images of the humans/objects, present them in real-time or at any distant location, and then transmit these images and corresponding voices/sounds to the receiver [21]. By this approach, holographic communication can ensure that the objects or the humans, with the real-time voice/audio information, present at a remote/different location to appear exactly the same at the location of investigation and it provides an experience much closer to reality, compared to Virtual Reality (VR) and Augments Reality (AR). Further, it has already been established that holographic communications and their variants will play a key role within the industry, agriculture, education, entertainment, and multiple other domains. As the major requirement, owing to real-time excessive data amount to be transmitted, the data rate required to stream the holographic media will be very large, requiring large throughput [10]. Also, networks will be needed to be reliable and secure to eliminate any jitter and guarantee low latency with privacy and security.

2) **Five-sense communications**: This will initiate a new form of remote human interactions [21], which will result in immersive five-sense communications. Such a five-sense medium will combine all the information related to sense organs of human, i.e., eyes, nose, ears, tongue and skin, Aforementioned technology will detect the change in the sensation in the body of human /environment and integrate the multiple sensed information via the neurological processes. Next, this data will be communicated to the distant location with the aid of the receiver, which will result in a completely immersive experience. Such communication will require the consideration of engineering and perceptual requirements during the design process to ensure that the service requirements can be determined [10].

3) **Wireless Brain-Computer Interfaces (WBCI)** will introduce interfaces that will utilize human thoughts for interaction with the machines and/or the environments [42]. WBCI initially reads the neural signals generated within the mind of any human being using a certain amount of electrodes. Then, it will translate the signals acquired into the commands that the machines understand, thereby ensuring that the control can be achieved [43]. It has been envisioned that along with tactile Internet, haptic communications, etc., WBCI will be a key use-case of the 6G technology [9]. In regard to the requirements, WBCI will need very high data rates, very low latency, and high reliability along with very high computing capabilities simultaneously guaranteeing the required Quality of Service (QoS) and Quality of Experience (QoE).

4) **Smart healthcare**: Existing studies in the literature have stated that the 6G technology is capable of

building smart healthcare systems wherein a reliable remote monitoring system, remote diagnosis, remote guidance, and remote surgery can be facilitated [44]. Owing to the high data rate, low latency, and ultra-reliability, the 6G technology can transport large data volumes faster with reliability. Further, with the aid of AI technology, the received data can be analysed efficiently by the doctors resulting inaccurate diagnoses. Using blockchain technology, personal data can be privately and safely shared, contributing to medicine development [10].

5) **Smart education**: Smart education will also be introduced by the 6G technology via innovative methods such as holographic communications, five-sense communications, mobile edge computing, and AI. These innovative methods will allow the learner to view multiple structures and models in the 3D format and help the instructor to deliver the content from remote locations leading to an interactive and immersive online education. Further, intelligent classes can be introduced wherein the data collected by the sensors will be sent to the cloud(s) or the edge cloud(s) for analysis. The collected data can then be used to analyze and improve education via students' feedback.

6) **Industry Internet**: In this scenario, the 6G technology will introduce multiple vertical industry types viz., electricity, manufacturing, delivery, ports, etc. With the help of AI, the 6G technology will automate the process control and operation of the systems/devices. Also, the decision-making and analysis through the data collected within the cloud/edge cloud will be intelligent and automated. Lastly, remote maintenance and control can be ensured through holographic communications, leading to cost reductions and safety [45]. Industry Internet will require rigorous low-latency, broadband, and reliable transmission.

7) **Fully autonomous/Self driving**: This can be obtained by the 6G technology via the deployment of multiple high definition cameras and high-precision radar sensors so that the vehicles can perform all the driving tasks, including the core functions of autonomous driving, which include perception, planning, and control [46]. The main challenge for achieving fully autonomous driving is the manner in which the stringent safety demands can be satisfied when scenarios with different driving conditions arise [10]. Further, the performance of autonomous driving can be improved using blockchain technology, and the complex information can be handled efficiently using AI and low-latency aspects of the 6G technology.

8) **Smart city**: In this scenario, the city will be able to operate in a smart and self-organized by collecting and analysing the huge data volumes obtained from multiple industries, from urban planning to garbage collection, thereby ensuring that public resources can be utilized efficiently, resulting in enhanced QoS for the citizens, and reduction of public administrations' costs [47]. The key challenge will be in regard to the connectivity and coverage capability of the 6G technology as multiple sensors, and intelligent terminals will be involved in this case [10].

9) **Intensive and Sensitive IoT**: IoT enabled by the 6G technology is anticipated to be highly intensive in terms of collected information and processing/computing the data, sensitive to privacy/security and associated delays [48]. The new and evolving drivers such as tactile Internet, holographic communications, five-sense communications, and WBCI will result in a surge of large volumes of data which will require large computation power for processing and analysis of data. Also, these will require stringent latency requirements and rigorous data security/privacy protection [10].

10) **IoT Evolution:** For 6G-IoT, collection of data, storing, query, understanding, and utilizing the raw sensor data will be more important. This will mandate 3C's 5G technology i.e., Communication, Caching, and Computing to converge to (4CSL) i.e., Communication, Computing, Caching, Control, Sensing, and Localization in the 6G technology. Also, the applications such as autonomous driving, remote control, localization, etc., require 4CSL for IoT-6G [49].

## 4. Use Cases

The technological drivers detailed in the previous section will aid in the realization of novel applications within the 6G-IoTs. In present section, we review the various evolving use cases related to 6G-IoTs.

1) **Healthcare IoT (HIoT)**: The opportunity of Healthcare Internet of Things (HIoT) is in the possible impending usage of new sensor and actuator technologies. For example, the use of non-invasive technologies in sensor measurement can offer more accurate and faster sensing of various critical body parameters. Availability of such conveniently deployable technologies (e.g., non-invasive sensing and imaging) can radically raise their usage in clinical settings. The evolution in the design of actuators can also advance the mechanization of routine medical tasks such as, drug delivery. The diagnostic systems involved in diagnosing vital organ diseases such as heart and brain promise to create more accurate devices using advancements in actuator technologies. Further, automation of the healthcare system can promise substantial improvements in patient health monitoring and hospital management. HIoT promises advancement in the procedures followed for routine measurements of patients, drug consumption, and the customized drug based on individual patient's health needs. On the technology end, HIoT offers most healthcare institutions an opportunity to get connected to the internet owing to the increasing accessibility of high bandwidth connectivity, inexpensive cloud storage and computation, and largescale data analytics. In this emerging scenario, HIoT technologies are important because of the personalization of clinical healthcare [50]. Further, the IoT revolution will occur through the enabling technologies of 6G-IoT.

The 6G technology enabled with cloud computing offers healthcare services with cloud storage, computing and analysis of biomedical data. The biomedical data obtained through advanced non-invasive sensors are stored to the cloud by optimizing constraints in communication resources and bandwidth. This brings the end user devices closer to the data source using edge computing technology. 6G will depend on the edge technology to provide the smooth and high speed internet services to the smart biomedical devices. Thus, important healthcare data is collected, processed and analysed in real time at the edge technology's nodes which are located near the medical sensors. For example, in cardiology, edge node will receive patients' electrocardiogram (ECG) data which is transmitted from non-invasive biomedical sensors and determines whether the patient is suffering from any cardiovascular disorder. The ECG signal data are recorded continuously and communicated to the edge nodes. These nodes process and analyse the biomedical data and communicates the vital information to the cloud for storage. 6G with cloud offers important advantages such as, reliability, low latency, scalability, privacy, and adaptability. The massive number of biomedical sensors from a hospital unit can connect to the 6G Internet; hence, all the benefits of edge computing in 6G will provide high QoS in healthcare. The authors in [51] have explained the possible utilization related to 6G technology by the illustration of mURLLC and THz communications, which will be used for supporting the medical field regarding healthcare related data transmission requiring very low latency, and providing impetus to the wearable technology within the market. Further, in [52], the authors have demonstrated a surgery method which can be conducted using the blockchain method. Also, doctors can observe and handle surgical procedures via video streaming on a real-time basis with the aid of medical robotics and biomedical devices that are connected through the 6G networks [44]. In the context of intelligence, AI/ML techniques must be exploited in the healthcare domain to ensure data learning. As an example, the authors in [53] have used ML methods such as, Bayesian Classifier, Logistic Regression, and Decision Tree for analysing the patients' medical history. The relevant data is collected through the wearable sensors. The authors introduce an inter-disciplinary approach by connecting the concepts of e-healthcare, priority, big data analytics and radio resource optimization by developing a 6G Heterogeneous Networks (HetNets). The authors have analysed stroke patients' medical data and recordings from non-invasive IoT medical sensors to predict the possibility of a future stroke [53]. In recent years, with the spread of the COVID-19 pandemic, major health concerns have been cited by various countries around the globe. In this regard, many aspects of computing at the edge, known as edge computing, and at cloud level, known as cloud computing; have been applied to combat COVID-19 in different manners [54]. Also, AI methods can be used for accurate and speedy disease diagnosis by processing the data using edge intelligence and computing techniques [55].

2) **Vehicular IoT (VIoT):** With the advances in the 6G technology, intelligent transportation systems are being revolutionized. In [56], the authors have exploited mMTC in VIoT networks enabled by the 6G technology for enabling the vehicle-to-everything connectivity in view of transmitting shorter vehicular information payloads through a large number of vehicles without any human interaction. The study clarifies that, in the future, within the 6G enable VIoTs, data rate prediction will be an open challenge for research owing to various complex interdependencies between different factors to manage the mobility, handle channels, and networking. Further, ML techniques provides throughput prediction with much efficiency for mimicking the attainable behaviours of the networks enabled via 6G by training the network with load information obtained by control channel analytics [57]. In order to leverage the potential intelligence within the VIoTs, combination of edge computing with Artificial Intelligence (AI) techniques can be implemented, to conduct the traffic volume estimation and forecast weather based on the data obtained [58]. Provisioning intelligence within the 6G enabled VIoTs through Deep Learning (DL) is studied in [59] wherein, optimality of the resource assignment algorithms on the basis of deep reinforcement learning has been studied to model the vehicular communication channel and for supporting the network management [60]. A similar focus on vehicular intelligence appears in [61]. The authors have adopted the DL techniques for autonomous data transmission scheduling via supervised, unsupervised, and reinforcement learning methods. The study also mentions that, in the future, cooperative DL approaches such as FL will be required to be developed since the 6G enabled VIoT networks will be highly scalable and distributed. Also, at the vehicular data controller, intelligent software based on DL is deployed in [62], thereby providing apt security solutions and protection in the 6G-VIoT networks. Further, regarding the Autonomous Vehicle (AV) applications, the 6G technology is envisioned to offer multiple opportunities for satisfying the required stringent service requirements to achieve reliability and high-speed communications [63]. For realizing AV in the 6G era, it will be mandatory to explore the performance vehicle-to-vehicle networks communication since every AV in the interconnected system will be recognized as an individual entity having complete control access.

3) **Unmanned Aerial Vehicles (UAVs):** Much research focus lately has been on the exploration of applications related to the 6G enabled UAV networks. The study in [64] has considered a cell-free UAV network for the 6G enabled wide-area IoT, focusing upon the UAV flight process-oriented optimization. The proposed method is supporting the massive access for wide-area IoT devices and is promising method in identifying the cell-free coverage patterns in the era of 6G era. The researchers in [65] have proposed IoT networks enabled by 6G with UAV-supported Clustered Non-Orthogonal Multiple Access (C-NOMA) method [66] to support the wireless powered communications. The authors in [67] have characterized the UAV to ground channel with arbitrary three-dimensional (3D) UAV trajectories for the UAV-based 6G networks, and the authors in [68] have studied a collaborative multi-UAV trajectory optimization and resource scheduling framework for a 6G-IoT network wherein, multiple UAVs fly as the base stations for transferring the energy to multiple terrestrial IoT users. Further, techniques such as AI are used to contribute optimal solutions to UAV networks enabled by 6G via edge intelligence techniques [69], and FL methods have been implemented in [70] and contributed to privacy preserved intelligence for the UAV-based 6G networks. As a major challenge, in future UAVs within the 6G-IoT networks, regulations are required to contribute and the necessary guidelines/regulations for deploying the UAVs within the IoT systems must ensure safety and privacy [71].

4) **Satellite Internet of Things (SIoT):** With the 6G technology, it will be mandatory to incorporate satellite communications with the existing deployed wireless networks in view of massive IoT coverage, which will result in the IoT [72]. With the advanced satellite technologies and 6G, multiple satellites are deployable at multiple orbits above the earth, and in this regard, the low earth orbit systems will be most suitable to realize the global coverage and reuse frequency efficiently [73]. In [74], the authors have focused the research considering a low earth orbit satellite network able to reinforce UAV trajectories' navigation to collect IoT data. The study in [75] provides further insights into the SIoTs as the authors have built a comprehensive model and have highlighted the relevant technical discussions and challenges. Further, as demonstrated in [76], RIS technology and satellite communications can be integrated to improve power of signal transmission strength at the satellites. The study in [77] has considered an energy-aware massive random-access scheme for satellite communications in the 6G-enabled global IoT. Further, spectrum sharing is another key issue to be addressed within the satellite communications for 6G-IoT networks [78].

5) **Industrial Internet of Things (IIoT):** Lately, 6G within the IIoT has been investigated. E.g., ML technology has been applied for providing intelligence within the 6G enabled IIoT networks [79]. Also, the transfer learning method is implemented in [80] for coordinating the distribution of data and transmission within a blockchain-enabled 6G IIoT network. The reasoning abilities of ML is employed in [81] to tackle the issues related to detection of fault and prediction of the faults in the aforementioned systems. Further, as highlighted in [82], AI techniques can be used for intelligent agriculture, and blockchain technology can enable secure production and logistics design. Lastly, even though 6G offers unprecedented benefits within the IIoT domain, it also presents the key challenges of security attacks and privacy issues are to be addressed. In regard to this, blockchain technology has already been identified as one of the probable way to solve the security and trust issues within the 6G enabled IIoT networks [84].

6) **Smart Agriculture:** In agricultural sector, smart farming is an optimum solution in order to meet the requirements of farmers in terms of food supply and demand along with safety and consistency. So the productivity/technology is enriched using specific environmental control and management by means of IoT devices like various sensors, their localization in robots and systems. Hence, resulting in the use of rarer resources and fewer waste compared to traditional farming. The role of edge computing in agricultural Sector is as follows: In agricultural sector, the customers are fundamentally the farmers and then agricultural industries, provided the solutions due to edge computing with the set of desired capabilities. Therefore, this use case provides a business platform for the farm owners to recognize the deployment of cost effective, increased crop yield etc., kind of smart agriculture. By means of 5G/B5G/6G, Multi-access Edge Computing (MEC), the challenges in smart agricultural systems such as, predictive analytics, weather forecasting and smart logistics and warehousing etc., will be addressed. The amalgamation of 5G/6G and edge computing in smart farming/agriculture is shown in Fig. 1.

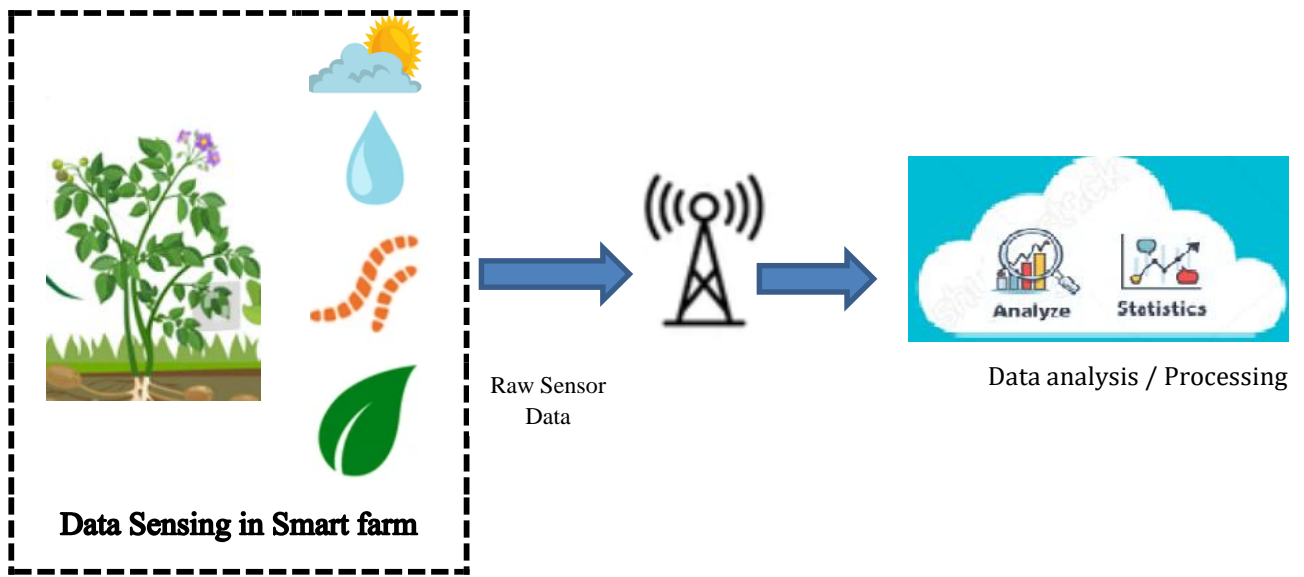

Fig. 1: MEC and 5G for Automated Crop Monitoring and management in Smart Farm.

The agriculture sector comprises of use cases like -
- **Sensing and monitoring:** Water consumption and utilization can be effectively monitored to confirm healthy crop yielding and minimal waste of water and related resources.
- **Data Analysis:** The remarkable advances in wireless sensor networks (WSNs) and 5G/B5G/6G communication technologies are going to provide the real-time information/data for the farmers about their land, crop etc., anytime and everywhere. The technology can also determine crop health status during various phases at regular intervals. Where 5G/B5G/6G are critical to precision agriculture using extensive Machine-to-machine (M2M) services. From weed monitoring, insect tracking etc., to greenhouse management, monitoring of climate conditions, Crop management, Cattle monitoring and management etc., overall in precision farming data analytics is required and hence can make use of edge computing.

TABLE 1: Use-cases of edge considerations for vertical industries.

| Edge Considerations | Agriculture | ITS | Manufacturing | Energy | Video Analytics |
|---|---|---|---|---|---|
| ✓ Architecture | Multi server | Multi server | Single server | Singe/Multi server | Single/Multi server |
| ✓ Edge location | RAN | RSU or RAN | On-premise | On-premise | 5G RAN |
| ✓ Storage capacity | Medium/High | Medium/High | Low | Low/Medium | Very High |
| ✓ Computing capacity | Medium | High | Low/Medium | Medium | Very High |
| ✓ Typical Latency Requirement | < 1 s | 10 ms - 1 s | < 100 ms | < 10 ms | < 1 s |
| ✓ Security Issues | DDoS and Data poisoning attacks | Data & ML model poisoning | Online adversarial attacks | Online adversarial and DDoS attacks | Data & ML model poisoning |
| ✓ Edge Intelligence | Anomalies detection, crop treatment recommendations | Object detection, tracking, traffic/vehicle prediction & control etc. | Anomaly detection, process automation & control etc. | Fault isolation, predictive maintenance etc. | Object detection, recommendation, summarization etc. |
| ✓ Real-world deployments | 5G-ENSURE | 5G-ENSURE | SONATA [4] | COGNET [5] | 5G-ENSURE [3] |

## 5. Open problems of research

In this section, we underline the various research challenges and mention the possible perspectives in the domain of 6G-IoT.

1. In regard to ML for massive 6G-IoT, the existing research studies have only provided an introductory perspective, and much research considering the key issues still remains to be conducted [85]. These challenges include the issue of the incomplete dataset are various layers, security, and privacy, long time cost, computation, and storage. Further, existing centralized techniques will not be applicable to the large-scale IoT challenges of 6G, and hence, ML techniques such as Distributed ML will be required, which needs much research attention. Lastly, considering the modifications in the ML techniques which will be required in the 6G-IoT, optimization of the new ML methods will be an open area for research.

2. Even though Blockchain technology is emerging and intended to play a crucial role in the 6G-IoT, many challenges need timely solutions before the implementation of Blockchain for the massive IoTs [86]. The first issue includes the performance of Blockchain-enabled systems, which needs to guarantee the simultaneous performance of both the overlaid and the underlaid layers within the IoT systems. A second issue is the blockchain technology requiring large amounts of computation, caching, and transmission to generate the block(s), verify, store the ledger, and consensus among the nodes. Specifically, the main advantage(s) of including Blockchain within the IoT requires investigation. The aforementioned also points towards the issue of allocating the resources between the Blockchain and the wireless system in view of ensuring the optimal utilization of the resources. Next, multiple security-related problems originate once large amounts of services are outsourced at the edge, which requires accurate investigation. Lastly, there are open problems for research within the domain of blockchain optimization. Specifically, an adaptive generic blockchain model for 6G-IoT needs to be developed to support multiple services. To meet the heterogeneous demands of the 6G-IoT, a novel consensus protocol must be designed for improving throughput, security, and privacy.

3. There are multiple issues regarding the security and privacy of 6G-IoTs, which require investigation and solutions [87]. This occurs mainly due to the integration of 6G within the IoT networks, which makes the 6G-IoT vulnerable to multiple threats that are associated to wireless interface attacks at computing units/servers due to unauthorized access of data, integrity threats within the access network infrastructure, and Denial of Service (DoS) to shut down the software and data centres.

4. A major concern in the 6G-IoTs will be achieving high energy efficiency since next-generation applications such as autonomous driving, space, terrestrial communications, etc., will require efficient network operations with efficient energy resources. Hence, much research is necessary to explore the issues of energy efficiency within 6G-IoT networks [88].

5. Another challenge within the 6G-IoTs is related to constraints faced in selection of IoT Devices hardware, because of the large communications and computations required in the 6G-IoTs [89]. Research to impart solutions based on targeted hardware such as hardware-based AI training solutions in the field of nano-IoT devices and in the field of enhancing services assisting the living such as the embedded wearables in 6G-IoT networks is much desired.

6. With the 6G technology integration within the IoT networks, there will be a transformation in the IoT markets, and the IoT ecosystems will be enabled via wireless systems. This will require the standardization of the stringent specifications through the collaboration of the various relevant stakeholders [90]. The role of network/system standards will be significant for deploying the 6G-IoT ecosystems on a extensive scale due to the dependence on several services including computing services and efficient communication protocols for 6G server-IoT device.